\documentclass[preprint,prb,amsmath,superscriptaddress,aps,subeqn,showpacs]{revtex4}
\usepackage{graphicx}
\usepackage{amssymb}
\usepackage{bm}
\usepackage{subfigure}
\usepackage{graphpap}
\usepackage{color}

\begin{document}

\title{Thermal transport properties of metal/MoS${_2}$ interfaces from first principles}

\author{Rui Mao}
\affiliation{Department of Electrical and Computer Engineering,
North Carolina State University, Raleigh, NC 27695-7911, USA}

\author{Byoung Don Kong}
\affiliation{Department of Electrical and Computer Engineering,
North Carolina State University, Raleigh, NC 27695-7911, USA}
\affiliation{Present address:  U.S. Naval Research Laboratory,
4555 Overlook Avenue, Washington, D.C., 20375, USA}

\author{Ki Wook Kim}\email{kwk@ncsu.edu}
\affiliation{Department of Electrical and Computer Engineering, North Carolina
State University, Raleigh, NC 27695-7911, USA}

\begin{abstract}

Thermal transport properties at the metal/MoS${_2}$ interfaces are analyzed
by using an atomistic phonon transport model based on the Landauer formalism and first-principles calculations.  The considered structures
include chemisorbed Sc(0001)/MoS${_2}$ and Ru(0001)/MoS${_2}$, physisorbed Au(111)/MoS${_2}$, as well as Pd(111)/MoS${_2}$ with intermediate characteristics.  Calculated results illustrate a distinctive dependence of thermal transfer on the details of interfacial microstructures.  More specifically, the chemisorbed case with a stronger bonding exhibits a generally smaller interfacial thermal resistance than the physisorbed. Comparison between metal/MoS${_2}$ and metal/graphene systems suggests that metal/MoS${_2}$ is significantly more resistive. Further examination of lattice dynamics identifies the presence of multiple distinct atomic planes and bonding patterns at the interface as the key origin of the observed large thermal resistance.
%the differences in mass, composition and interaction strength as the main $mechanisms for the large interfacial thermal resistances of metal/MoS$_2$ $systems.
%Extention calculation of other metal/TMD contacts also confirms the above %observations.
%The result clearly suggests the feasibility
%of phonon engineering for thermal property optimization at the interface.

\end{abstract}


\maketitle

\section{Introduction}

Transition-metal dichalcogenides (TMDs) have emerged as key candidates for the beyond-graphene, two-dimensional (2D) or van der Waals crystals due to their distinctive electrical, optical, and thermal properties.~\cite{xu2013, mak2010}  In each case, the bulk material is formed by a stack of 2D monolayers through the weak van der Waals interactions as in graphene, while the intralayer binding is much stronger.  For instance, monolayer of molybdenum disulfide (MoS${_2}$)$-$a prototypical example$-$consists of one Mo plane sandwiched between two S planes via the covalent bonding that is arranged in a trigonal prismatic network.~\cite{mak2010}  Consistent with its diatomic nature, MoS${_2}$ exhibits a non-zero energy gap whose magnitude depends on the layer thickness.~\cite{mak2010,kam1982,han2011}  Successful fabrication of a transistor with a large on/off ratio (as high as $ 10^{8}$ owing to the large gap)~\cite{radisavljevic2011} has made this material an early focus of investigation among the TMDs.

As in the metal/graphene (metal/Gr) cases, it was recently found that the metal/MoS${_2}$ interfaces can be classified into two categories$-$physisorption and chemisorption.~\cite{gong2010,popov2012} The former generally has a smaller binding energy and a larger interfacial separation than the latter.
Since the contacts with metallic electrodes comprise a crucial component of the  high-performance electronic devices, considerable efforts have been devoted to investigate the electrical transport properties of the metal/MoS${_2}$ structures.~\cite{kang2012,gong2013}
In comparison, thermal transport has received much less attention. Nonetheless, the impact of efficient heat transfer on the operation of 2D crystal devices is significant when considering the inevitable presence of heterogeneous interfaces  and subsequent joule heating in the layered structures.~\cite{chiritescu2007,muratore2013,Li2011}  Accordingly, a comprehensive understanding of thermal properties at the interface with the metallic contacts is crucial from the perspective of both fundamental low-dimensional physics and practical applications of this emerging material system.

In this paper, we present a detailed theoretical analysis of interfacial thermal resistance in the metal/MoS${_2}$ system via phonon transport. The sample structures are chosen to reflect the range of typical interfaces from chemisoprtion to physisorption. For an atomistic description of lattice dynamics at the interface, our theoretical approach adopts a first-principles method based on density functional theory (DFT) and density functional perturbation theory (DFPT).~\cite{baroni2001,gonze1997}  Then the phonon/thermal transport characteristics are determined via the Green's function techniques~\cite{lee1981,datta1997,nardelli1999} in the Landauer formalism.~\cite{landauer1970}    The calculation results are examined in terms of interfacial microstructures and force constants to identify the key contributors to the disparate thermal properties at the considered interfaces.  Comparison is also made with those of the corresponding graphene based structures.~\cite{mao2013}

\section{Theoretical Model}
Thermal conduction across the heterogeneous metal/MoS${_2}$ interface is characterized by the interfacial resistance or the so-called Kapitza resistance.~\cite{pollack1969,swartz1989}
In the phonon transport calculation, we consider a three parted system where the central interface region (i.e., the region of interest) is connected to the thermal reservoirs on the left and the right with two semi-infinite leads (labeled $L$ and $R$), often known as the lead-conductor-lead configuration.~\cite{nardelli1999,zhang2007}
In the nanoscale, the Kapitza resistance can be evaluated by extending the Landauer formalism for electrons to phonons. Then, the thermal current density can be written as
\begin{equation}
\label{thermalcurrent}
J(T_L,T_R)=\frac{\hbar}{2\pi}
\int_0^{+\infty}{d\omega}~\omega~
\mathcal{T}_{ph}(\omega) \left[ n(T_L,\omega) - n(T_R,\omega)\right],
\end{equation}
where $n(T_{L,R},\omega)$ is the equilibrium Bose-Einstein distribution for phonons,  $T_{L,R} = T\pm\Delta T/2$ is the temperature in the left or right thermal reservoir, and $\mathcal{T}_{ph}(\omega)$ is the phonon transmission
function through the structure.  In the limit of small $\Delta T$, the phonon contribution to the thermal conductance $\kappa_{ph}(T)=J(T)/\Delta T$ is then given by
\begin{equation}
\label{thermalconductance}
\kappa_{ph}(T)=\frac{1}{{2\pi}\hbar}\int_0^{+\infty}{d(\hbar\omega)}~ \mathcal{T}_{ph}(\omega)~\hbar\omega\left[\frac{\partial n(T,\omega)}{\partial{T}}\right].
\end{equation}
The thermal resistance (normalized to the cross-sectional area of the interface) is then obtained by inverting Eq.~(2).
The phonon transmission function $\mathcal{T}_{ph}(\omega)$ can be
calculated by adopting a real-space Green's function approach, similar to the one used for electronic transport.~\cite{nardelli1999} In particular, one can take advantage of the following analogy between the electronic and phononic systems:
$
E_{el}{I}\leftrightarrow \omega^2{M}_{ph}
$
and
$
{H}_{el}\leftrightarrow {K}_{ph}.
$
Here ${H}_{el}$ and $E_{el}$ are the Hamiltonian and the eigen-energy in the electronic system, whereas ${K}_{ph}$ and ${M}_{ph}$ denote the matrix of the interatomic force constants (IFCs) and the diagonal matrix corresponding to the mass of the atoms.~\cite{gonze1997,calzolari2012} Additionally, $I$ symbolizes the identity matrix and $\omega$ the phonon frequency.  Further details on the theoretical formulation can be found in Refs.~\onlinecite{calzolari2012} and \onlinecite{mao2013}.

%\section{Numerical implementation}

In the present treatment, the IFCs (thus, ${K}_{ph}$) are calculated fully from the first principles within the DFT/DFPT framework that allows accurate consideration of the microscopic geometry as well as the chemical and electronic modification at the interface without resorting to phenomenological or {\it ad hoc} models.~\cite{baroni2001,calzolari2012}
Specifically, the QUANTUM-ESPRESSO package~\cite{giannozzi2009} is used with ultrasoft pseudopotentials in the generalized gradient approximation (GGA). A semiempirical van der Waals force correction is also added to the density functional calculation (GGA+D) to obtain more accurate interlayer distances.~\cite{grimme2006,barone2009}  It has been verified that the GGA+D routine provides the optimal results for layered MoS$_2$ structures in comparison to other approaches such as the nonlocal exchange-correlation functional treatment and the local density approximation.~\cite{ataca2011}  A minimum of 50 Ry is used for the energy cut-off in the plane wave expansion along with the charge truncation of
600 Ry. In addition, the Methfessel-Paxton first-order spreading with the smearing width of 0.01~eV is employed.  The momentum space is sampled on a 6$\times$6$\times$2 Monkhorst-Pack mesh in the first Brillouin zone.
%Up to six layers of metal atoms are included in the interface region,
%which is tested to be suffice to recover the bulk metal characteristics. A minimum vacuum region of 15 %{\AA} is used to ensure that there is no coupling between neighboring slabs.
The realistic interface structures are obtained through geometry optimization, where the total energy and atomic force are minimized.
The energy convergence threshold is chosen at $10^{-8}$ Ry and the maximum forces acting on each atom is relaxed below $10^{-4}$ Ry.

\section{Results and Discussion}

In order to achieve the maximum orbital overlap (i.e., a good electrical contact), it is highly desirable to minimize the lattice constant mismatch between the metallic material and MoS${_2}$.  In addition, the work function of metal species must be close to that of the conduction band minimum or the valence band maximum for a small Schottky barrier, although the Fermi-level pinning may affect the final barrier height.~\cite{chen2013,das2012}
Following these criteria, the metal/MoS${_2}$ structures selected for the current investigation are  Au(111)/MoS${_2}$, Pd(111)/MoS${_2}$, Ru(0001)/MoS${_2}$ and Sc(0001)/MoS${_2}$.
The lattice constant of MoS${_2}$
is fixed at the optimized value 3.22 {\AA}. The 1$\times$1 unit cell
of face-centered cubic Sc(0001) is commensurate with
the 1$\times$1 MoS${_2}$ with only a 2.4\% lattice mismatch,
whereas the 2$\times$2 unit cell is needed for Au(111), Pd(111) and Ru(0001)
to make the lattice mismatch below 3.4\% against the
$\sqrt{3}\times\sqrt{3}$ R30$^\circ$ unit cell of MoS${_2}$.
Figure~1 shows the resulting interfacial structures of the
considered material combinations after geometry optimization.
In accord with earlier studies,~\cite{popov2012,chen2013} our calculation clearly
illustrates that Ru and Sc form strong bonding with MoS${_2}$ at the interface
(i.e., chemisorption) resulting in a relatively small interfacial separation (2.20 {\AA}
and 2.02 {\AA}, respectively).  On the other hand,  Au is physisorbed on MoS${_2}$
through weak van der Waals bonding with a larger interlayer distance (2.77 {\AA}).
As for Pd(111)/MoS${_2}$, the interfacial separation of 2.18 {\AA} is obtained
in close agreement with a recent DFT calculation (weak chemisorption).~\cite{chen2013}
These interfacial structures serve as the central region in the previously
mentioned lead-conductor-lead configuration. Two leads consisting
of respective bulk materials (i.e., bulk metal and bulk MoS$_2$) are connected seamlessly
to the interface region and modeled separately.  No appreciable mismatch
(i.e., resistance) exists between the leads and the conductor.

%\subsection{Interfacial Thermal Resistance}
Phonon transport through the different metal/MoS$_2$ interfaces is calculated in Fig.~2 by
using the theoretical model described earlier.   The results are plotted only up
to 100 cm$^{-1}$ in order to illustrate clearly the contributions of dominant low-lying acoustic
branches.  The impact of high-frequency optical phonons is negligible while not shown
explicitly. As evident from the figure, the transmission function of the physisorbed
Au/MoS$_2$ interface exhibits more resonant features than those of the chemisorbed cases.
This is due to the fact that the Au and S atoms are bonded through the weak van der Waals
force, leading to limited hybridization of vibrational modes at the interface. Accordingly,
phonon transmission is more selective with certain frequencies blocked almost completely (i.e.,
nearly zero transmission).  On the other hand, the strong interactions at the chemisorbed
interfaces (e.g., Sc/MoS$_2$ and Ru/MoS$_2$) result in substantially mixed properties
between the corresponding metal and MoS${_2}$.  Hence, phonon propagation
encounters a more gradual
barrier with much less hindrance over a broader frequency spectrum [see Figs.~2(b) and 2(c)].
As for the Pd/MoS${_2}$ contact, the transmission function in Fig.~2(d) resembles those of
chemisorbed cases although to a lower degree.  The bonding between the Pd and S atoms
appears to be not as strong as the other two cases, particularly the Sc/MoS${_2}$
structure.  Such intermediate characteristics were also observed when Pd is paired with
graphene.~\cite{mao2013}  The Pd/Gr interface was deemed a mixture of chemisorption and
physisorption with only weak, incomplete hybridization.

Figure~3 shows the interfacial thermal resistance obtained as a function of temperature.
Since the total resistance of the structure  contains the contribution from the leads
as well, the intrinsic thermal resistance at the junction is deduced by subtracting this
portion in a manner analogous to electrical transport.~\cite{datta1997}
The results exhibit the 1/T dependence in the low temperature region, while staying almost
invariant between 200 K and 450 K.
The dashed vertical line marks the values at room temperature, which are
$5.8 \times 10^{-8}$ Km$^2$/W, $1.9 \times 10^{-8}$ Km$^2$/W, $3.1 \times 10^{-8}$ Km$^2$/W,
and $1.2 \times 10^{-7}$ Km$^2$/W  for Au/MoS${_2}$, Sc/MoS${_2}$, Ru/MoS${_2}$, and Pd/MoS${_2}$,
respectively.  Consistent with the expectation from the transmission function comparison,
the chemisorbed interfaces (Sc/MoS${_2}$ and Ru/MoS${_2}$) show the lowest thermal resistances
among the considered.  Of the two chemisorbed examples, Sc/MoS$_2$ provides a smaller value than Ru/MoS$_2$
that can be understood, in part, by the difference in the interfacial separation (2.02 {\AA} versus 2.20 {\AA}) as the interatomic
distance tends to indicate the strength of the bonding between the atoms.  In this regard,
Pd/MoS${_2}$ provides an exception with the largest resistance even though it is
supposed to be chemisorption albeit weakly.  The physisorbed Au/MoS$_2$ is actually placed in
between the Pd/MoS${_2}$ and the (strongly) chemisorbed cases.  A similar feature was reported in
the Pd/Gr structure earlier.~\cite{mao2013}

%is 2.02 {\AA}  at the interface, which is smaller than the 2.2 {\AA} of Ru/MoS$_2$ contact. This observation becomes more clear during  the interatomic force constant analysis, which is carried out later in this section.
%While the Pd/MoS${_2}$ offers a $R_{in}$ larger than that of the physisorbed Au/MoS${_2}$, which may be attributed to the intermediate interaction nature of Pd contact with 2D materials.,ran2009}
%The above results of metal/MoS${_2}$ show good agreement with the experimental measured data of cross-plane thermal resistance of thin-film MoS${_2}$ ($ \sim 3.33-6.67 \times 10^{-8}$ Km$^2$/W).~\cite{muratore2013}
%Their measurement results of MoS${_2}$ includes the turbostratic stacking effect and defect scattering, Therefore, it is even higher than those of chemisorbed Sc/MoS${_2}$ and Ru/MoS${_2}$.
%This may imply a new avenue of tuning interfacial thermal resistance through misorientation, which is reserved for future investigations.  A more detailed discussion of IFCs can be found later in this section.

Considering their seeming resemblance, a detailed comparison between the metal/MoS$_2$ and metal/Gr systems could provide an insight into the lattice dynamics at the 2D crystal heterojunctions. The most crucial finding is that the metal/MoS$_2$ interfaces exhibit considerably larger resistances than the metal/Gr counterparts.  For instance, Ni/Gr that is a typical chemisorbed metal/Gr interface can reach a low thermal resistance of $3.9 \times 10^{-9}$ Km$^2$/W, while the lowest value for the chemisorbed metal/MoS$_2$ is about five times higher at $1.9 \times 10^{-8}$ Km$^2$/W.
This difference suggests strong dependence of the Kapitza resistance on the specifics of the interfacial microstructures.
To further illustrate this point, two essential factors$-$the atomic scale morphology and IFCs$-$are carefully examined at the boundaries.

The schematics in Fig.~4 highlights the dissimilarity in the interfacial structures of Ni/Gr and Ru/MoS$_2$. Unlike graphene that is formed by a single plane of C atoms, each layer of MoS$_2$ consists of one monatomic Mo plane sandwiched between two monatomic S planes. This multi-layer structure of MoS$_2$ damps the phonon vibrations across the interface.
Additionally, the Mo and S atoms are much heavier than C atoms (S-32.065 u, Mo-95.96 u vs.\ C-12.01 u in the unified atomic mass unit), which also contributes significantly to the reduced phonon transfer.  In fact, experimental investigations have observed the low thermal conductivity in both bulk and thin-film TMDs due to the high average mass, atomic complexity and weak bonding.~\cite{muratore2013}
A closer scrutiny shows that the larger mass variation may be yet another reason for the larger thermal resistance of metal/MoS$_2$.
When phonons propagate through the interface region, they encounter a drastic change in the atomic mass. In the case of Ru/MoS$_2$, for instance, it varies from 101.07 u to 32.065 u then to 95.96 u or vice versa. In other words, the relatively lighter S atoms sandwiched between the heavy Mo and contact metal atoms serve as an extra scattering layer due to the mass disorder.  In this regard, the case of metal-S-Mo$_2$ is analogous to the hydrogen terminated SiC on graphene (SiC-H/Gr), where the H adatoms provides an additional scattering mechanism.~\cite{mao2012}
On the other hand, the conditions are much more straightforward in the metal/Gr cases, with only the metal/carbon interaction at the interface; the phonons experience only a single alteration in terms of mass (from 58.69 u to 12.01 u for Ni/Gr).

Analyzing the impact of the second factor (i.e., the IFCs), Fig.~5 provides the interlayer force constants for the two chemisorbed cases, Ru/MoS$_2$ and Ni/MoS$_2$, deduced from the DFPT calculation.  The height of each bar symbolizes the interaction strength between two neighboring layers.  For example, the first bar on the left denotes the interaction between layers 1 and 2; the next bars are for layers 2 and 3, and so on.  In both of these plots, the metallic layers are up to layer 5. For MoS$_2$, layers 6 to 8 correspond to the covalently bonded S-Mo-S planes (i.e., the first monolayer from the interface). Accordingly, the S atoms in layer 9 belong to the second monolayer of MoS$_2$.  On the other hand, layers 6 to 10 represent the first through fifth graphene layers from the interface that are held together by the van der Waals force.
We focus on the force constants between layers 5 and 6, where the physical interface of two heterogeneous materials is located.  The magnitudes of these force constants indicate the interaction strength between the metal atoms and either the S or C atoms.
As shown, the force constant between Ru and MoS$_2$ is around 0.03 a.u.\ that is much smaller than the corresponding quantity of approx.\ 0.1 a.u.\ between  Ni and graphene (where a.u.\ denotes atomic Rydberg units).  The suggested weaker interaction between the metal and the S atoms is further verified by the analysis of electronic binding energy for the chemisorbed interfaces available in the literature.~\cite{popov2012,gong2013}  Clearly it is not unreasonable to anticipate lower phonon/thermal transmission at an interface with the less effective bonding and the more complex morphology.

In the case with Au or Pd, the physical picture appears to be somewhat different. Our calculation as summarized in Table~I indicates that the Au/MoS$_2$ and Pd/MoS$_2$ contacts exhibit interfacial force constants similar to the corresponding Au/Gr and Pd/Gr cases.  In fact, those with MoS$_2$ are slightly larger than the graphene counterparts.~\cite{popov2012,gong2010}  With the binding interaction much weaker than the chemisorbed, the distinguishing factor for the thermal resistance at these interfaces may be the mass variation/disorder rather than the magnitude of force constant.  Accordingly, MoS$_2$ again shows a larger resistance than graphene when interfaced with Au or Pd.  Nonetheless, the relatively muted differences between metal/MoS$_2$ and metal/Gr in the physisorbed (and the intermediate) cases can be attributed to the comparable bonding strengths.

\section{Summary}

Thermal transport in the metal/MoS$_2$ heterostructures is investigated by using an atomistic model based on the DFT formalism and the Green's function approach.  The obtained characteristics indicate generally more effective thermal transfer at the chemisorbed surface owing to the stronger interaction with MoS$_2$.  One exception is Pd/MoS$_2$ with a hybrid bonding at the interface that actually shows the largest interfacial thermal resistance among the considered. Comparison with metal/Gr reveals that metal/MoS$_2$ interfaces are more
resistive in terms of phonon/thermal transport.  A detailed examination of interfacial geometry and the lattice dynamics identifies the difference in atomic scale morphology, composition and interaction strength as the main origin of resistive nature in the metal/MoS$_2$ system. More specifically, the three-plane structure with heavy atoms, the mass disorder introduced by the light-massed sulfur plane as well as the different bonding forces at the interface, all contribute to phonon scattering and subsequently a large interfacial thermal resistance.  As these features are not unique to MoS$_2$, other TMDs are expected to be similarly resistive in heat transfer when interfaced with a metal.
%Further analysis of other metal/TMDC contacts also confirms these observations as the main mechanisms for larger interfacial thermal resistance

\begin{acknowledgments}
The authors would like to thank Cheng Gong for useful discussions.  This work was supported, in part, by SRC/NRI SWAN.
\end{acknowledgments}

\clearpage
\begin{table}
\caption{Thermal properties at the relevant metal/MoS$_2$ and metal/Gr interfaces. a.u. denotes atomic Rydberg units.}
\setlength{\tabcolsep}{12pt}
\centering
    \begin{tabular}{c c c c c} %{c c c c{5cm}}
        \hline\hline
        ~      & Bonding         & Interfacial       &Interfacial force       & Thermal resistance
\vspace{-8pt}	\\
		   & characteristics		   & separation	({\AA}) &constant (a.u.) &  (Km$^2$/W)              \\ \hline %[0.5ex]%
	Sc/MoS$_2$  & Chemisorption    & 2.02        & 0.043        & $1.9 \times 10^{-8}$                           	 \\
	Ru/MoS$_2$  & Chemisorption    & 2.20         & 0.034        & $3.1 \times 10^{-8}$                          	 \\
	Au/MoS$_2$  & Physisorption    & 2.77         & 0.005        & $5.8 \times 10^{-8}$                          	 \\
	Pd/MoS$_2$  & Mixed            & 2.18        & 0.0136       & $1.2 \times 10^{-7}$                          	 \\							
    Ni/Gr$^a$  & Chemisorption     & 2.02        & 0.103        & $3.9 \times 10^{-9}$                          	 \\
    Au/Gr$^a$  & Physisorption     & 3.31        & 0.004        & $1.7 \times 10^{-8}$                         	 \\
    Pd/Gr$^a$  & Mixed             & 2.43        & 0.0125        & $3.4 \times 10^{-8}$                         	 \\
        %[1ex]%
        \hline\hline
$^a$Ref.~\onlinecite{mao2013} & &
        \vspace{-8pt}\\
\end{tabular}

\label{Table}
\end{table}

\clearpage \noindent Figure Captions

\vspace{0.5cm} \noindent Figure~1. (Color online)
Side view of the metal/MoS$_2$ systems under consideration: (a) Sc(0001)/MoS$_2$, (b) Ru(0001)/MoS$_2$, (c) Pd(111)/MoS$_2$, and (d) Au(111)/MoS$_2$.
Two upper layers represent MoS$_2$.

\vspace{0.5cm} \noindent Figure~2. (Color online)
Phonon transmission function vs.\ frequency at the interface in the metal/MoS$_2$ structures under consideration. The magnitude of the transmission function can be larger than 1 since it also reflects the number of available modes.

\vspace{0.5cm} \noindent Figure~3. (Color online)
Interfacial thermal resistances vs.\ temperature for (a) Au/MoS$_2$, (b) Sc/MoS$_2$ (c) Ru/MoS$_2$ and (d) Pd/MoS$_2$. The vertical dashed lines mark the resistances at room temperature (300 K).

\vspace{0.5cm} \noindent Figure~4. (Color online)
Schematic view of the interfacial nanostructure
for Ni/Gr and Ru/MoS$_2$ with the atomic mass listed for the compositional atoms (in the unified atomic mass unit u).
The red solid box highlights the difference between graphene and MoS$_2$
in the atomic configuration and morphological arrangement. The dashed blue box shows
the mass disorder introduced by the sulfur layer.

\vspace{0.5cm} \noindent Figure~5. (Color online)
Interlayer force constants for Ru/MoS$_2$ and Ni/Gr.  The height of each bar represents the interaction strength between two layers, where a.u. stands for atomic Rydberg units.  In both plots, the metallic layers are up to layer 5 (i.e., 1$-$5).  The dash-dotted line indicates the physical interface with the metal as MoS$_2$ (in the S-Mo-S order) or graphene starts from layer 6.

%\clearpage
%\begin{center}
%\begin{figure}
%\includegraphics[scale=1.0]{metal-mos2_f1.eps} %[bb=16 388 543 684, width=8.5cm]
%\caption{Mao et al.}
%\label{Structure}
%\end{figure}
%\end{center}

%\clearpage
%\begin{center}
%\begin{figure}
%\includegraphics[scale=1.0]{metal-mos2_f2.eps} %bb=0 13 240 174,
%\caption{Mao et al.}
%%\label{Transmission}
%\end{figure}
%\end{center}

%\clearpage
%\begin{center}
%\begin{figure}
%\includegraphics[scale=1.0]{metal-mos2_f3.eps}
%\caption{Mao et al.}
%\label{Rin}
%\end{figure}
%\end{center}

%\clearpage
%\begin{center}
%\begin{figure}
%\includegraphics[scale=1.0]{metal-mos2_f4.eps}
%\caption{Mao et al.}
%\label{Microstructure}
%\end{figure}
%\end{center}

%\clearpage
%\begin{center}
%\begin{figure}
%\includegraphics[scale=1.0]{metal-mos2_f5.eps}
%\caption{Mao et al.}
%\label{Forceconstant}
%\end{figure}
%\end{center}

\end{document}